\documentclass[]{IEEEphot}

\usepackage{subfig}
\usepackage{caption}
\usepackage[font=scriptsize]{caption}

\jvol{xx}
\jnum{xx}
\jmonth{XX}
\pubyear{XX}

\begin{document}

\title{Beam Manipulation by Hybrid Plasmonic-Dielectric Metasurfaces}

\author{
	~K.~Arik$^1$ 
}

\affil{
	   $^1$	School of Electrical Engineering,                       Sharif University of Technology\\} 


\maketitle

\markboth{IEEE Photonics Journal}{Beam Manipulation by Hybrid Plasmonic-Dielectric Metasurfaces}


\begin{abstract}
A hybrid plasmonic-dielectric metasurface is proposed in order to manipulate beam propagation in desired manners. The metasurface is composed of patterned hybrid graphene-silicon nano-disks deposited on a low-index substrate, namely silica. It is shown that the proposed hybrid metasurface simultaneously benefits from the advantages of graphene-based metasurfaces and dielectric ones. Specially, we show that the proposed hybrid metasurface not only provides reconfigurability, just like previously proposed graphene-based metasurfaces, but also similar to dielectric metasurfaces it is low loss and CMOS-compatible. Such exceptional features give the metasurface exceptional potentials to realize high efficient optical components. To demonstrate the latter point, focusing and anomalous reflection are performed making use of the proposed hybrid structure as examples of two well-known optics functionalities. This work opens up a new route in realization of reconfigurable meta-devices with widely real-world applications which cannot be achieved with their passive counterparts.
\end{abstract}


\section{Introduction}

Wavefront shaping of light has been of utmost increasing attention since the very beginning time, as for example, lenses and prisms have been traditionally employed to focus and refract the light \cite{1}. These conventional optical devices, which are indeed based on phase accumulation along the light propagation path, are suffering from the bulky geometry as well as their relatively high reflection loss, restricting their application in the planar configurations \cite{2,3,4}.\\
\indent The attempts to overcome the aforementioned restrictions of the  conventional optical devices have stimulated the concept of frequency selective surfaces (FSS) \cite{5}. Such appealing surfaces, which are usually composed of planar arranged metallic patches or apertures, have the potential to control either reflection or transmission of electromagnetic waves over a highly thin surface. This can be accomplished by properly tuning the elements' size, shape, and thickness of the FSS \cite{6}. \\
\indent In a different area of research, the attempts to improve the characteristics of the conventional optical components has inspired introducing artificial engineered composite materials, known as man-made materials or "metamaterial". These engineered materials have attracted a lot of attention from physicist to realize novel and interesting optical components due to the fact that they provide a new degree of freedom to control electromagnetic waves in desired manners \cite{7,8,24,28,30}. However, their fabrication is still a challenge for the current technology due to their bulky geometry \cite{7,32}.\\
\indent Very recently, by combining the concepts of metamaterilas and frequency selective surfaces, two dimensional metamaterials, also known as metasurfaces, have been proposed. Just the same as metamaterials, metasurfaces provide an exceptional ability to control electromagnetic waves in unusual ways, while at the same time their FSS-like planar configuration eases the fabrication process \cite{9,11}. It has been  demonstrated recently that by imparting a controlled gradient of phase discontinuity over the surfaces, the reflected or transmitted field of the metasurface can be manipulated to the desired directions, apart from those expected by conventional Snell's laws of refraction \cite{12}. \\
\indent There are basically two types of metasurfaces proposed so far to manipulate electromagnetic waves: plasmonic metasurfaces among which those based on graphne plasmonics has been of most interest \cite{3,11,13,31}, and dielectric metasurfaces \cite{14,15,16,26,27,33}. Each of these structures has its own pros and cons. In particular, while plasmonic metasurfaces, more specifically graphene-based ones, provide excellent features such as high confinement of the field and reconfiguration, they suffer from high values of ohmic losses \cite{3}. In reverse, all-dielectric metasurfaces offer lower amount of non-radiation losses as well as compatibility with CMOS technology, while they can not be reconfigured \cite{15}.\\
\indent In this contribution, we propose a hybrid plasmonic-dielectric structure in order to simultaneously benefit from the advantages of plasmonic and dielectric metasurfaces. We will show that our proposed hybrid metasurface not only provides reconfigurability, but also is low loss and CMOS-compatible. Such exceptional features enable us to realize efficient optical devices. To illustrate the latter, two optics functionalities, namely focusing and anomalous refraction are performed making use of the proposed hybrid structure.
\begin{figure}
	\centering
	\subfloat[\label{1a}]{%
		\includegraphics[width=0.5\textwidth]{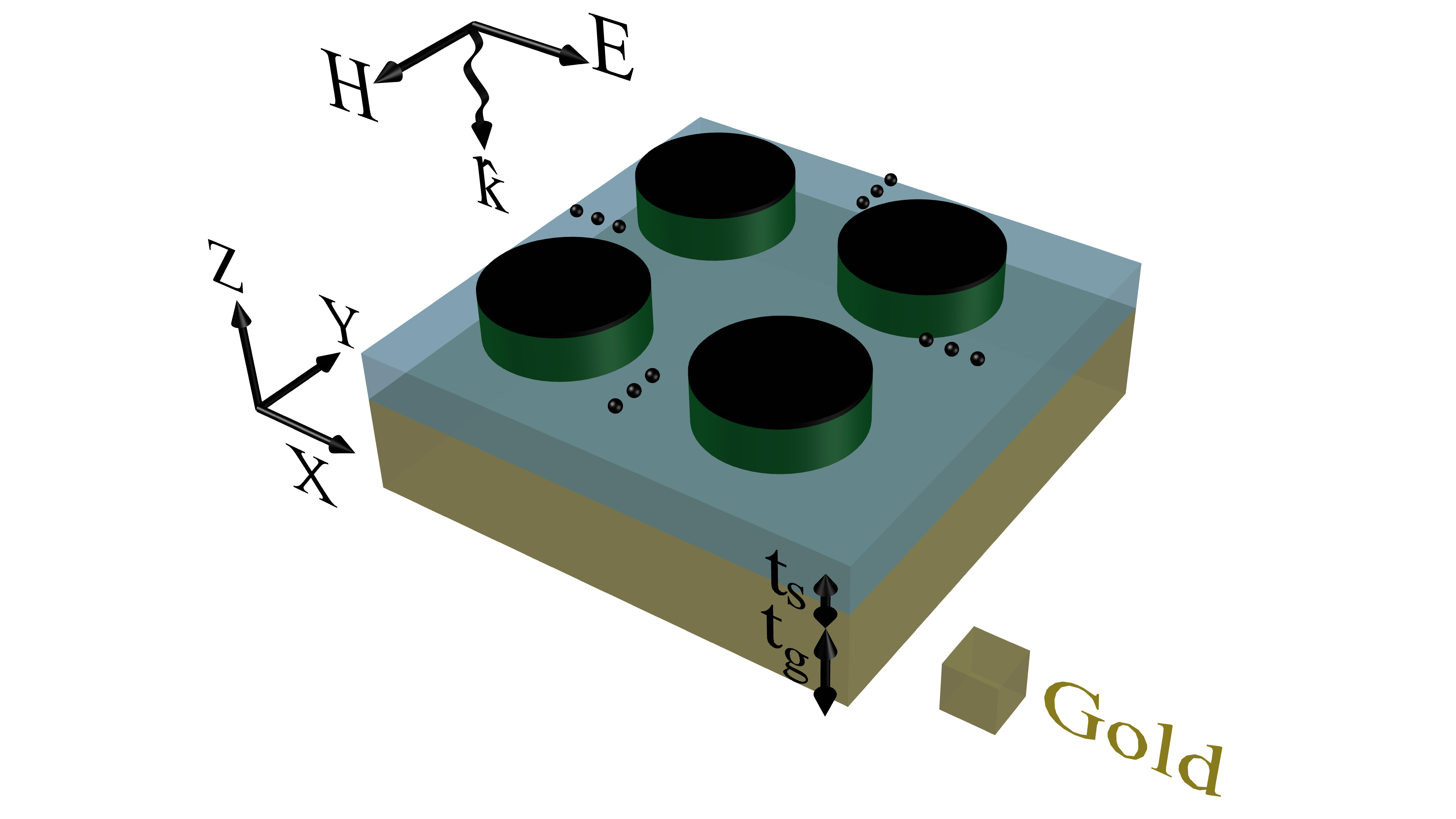}
	}
	\subfloat[\label{1b}]{%
		\includegraphics[width=0.5\textwidth]{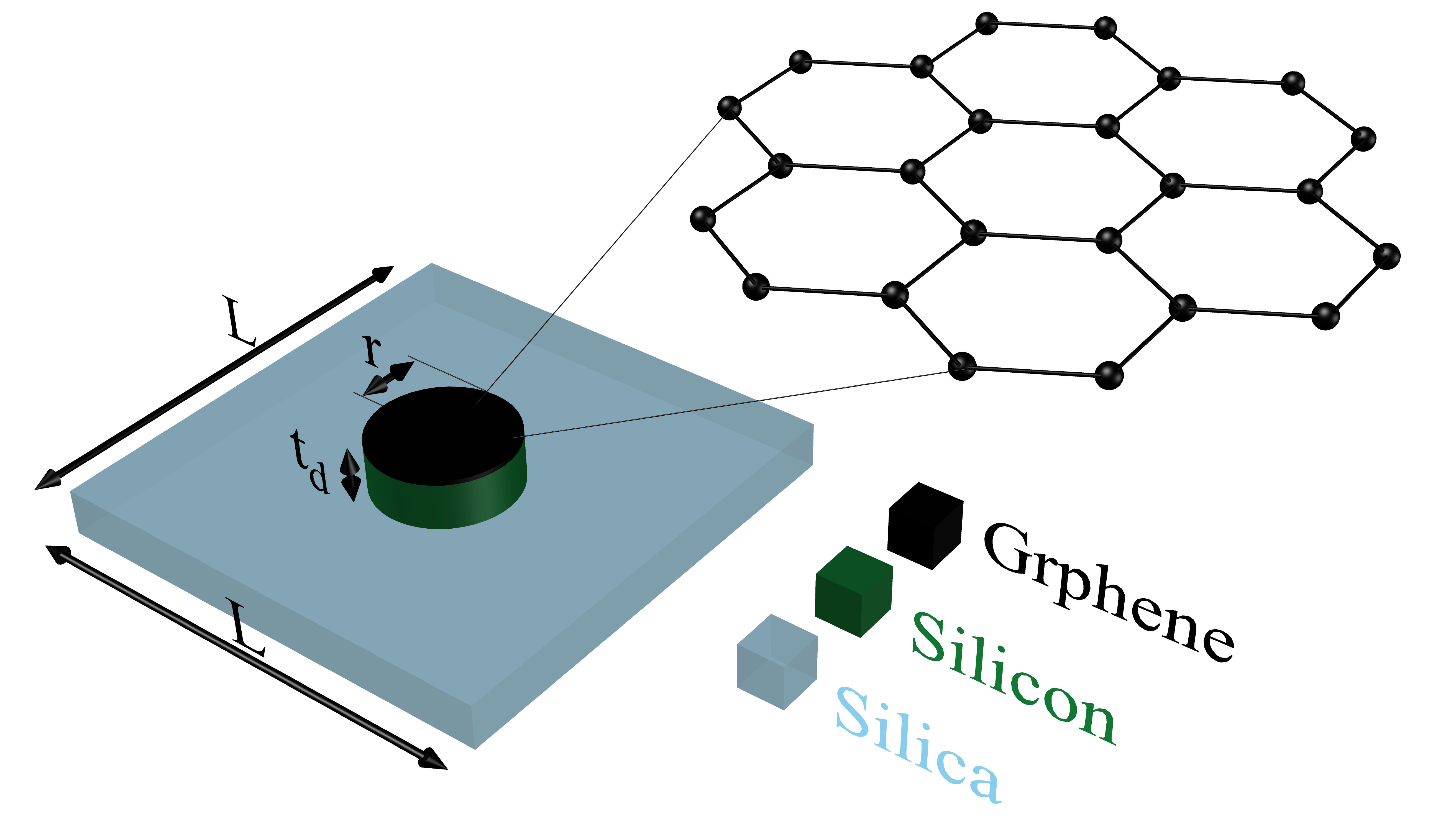}
	}
	\caption{(a) Schematic of the proposed metasurface for beam manipulation: a hybrid layer of silicon and graphene is deposited on a silica substrate placed on top of a thick gold metallic layer, (b) schematic of a single unit-cell of the proposed hybrid metasurfce.}
	\label{fig1}
\end{figure}
\begin{figure}[b]
	\centering
	\subfloat[\label{2a}]{%
		\includegraphics[width=0.3\textwidth]{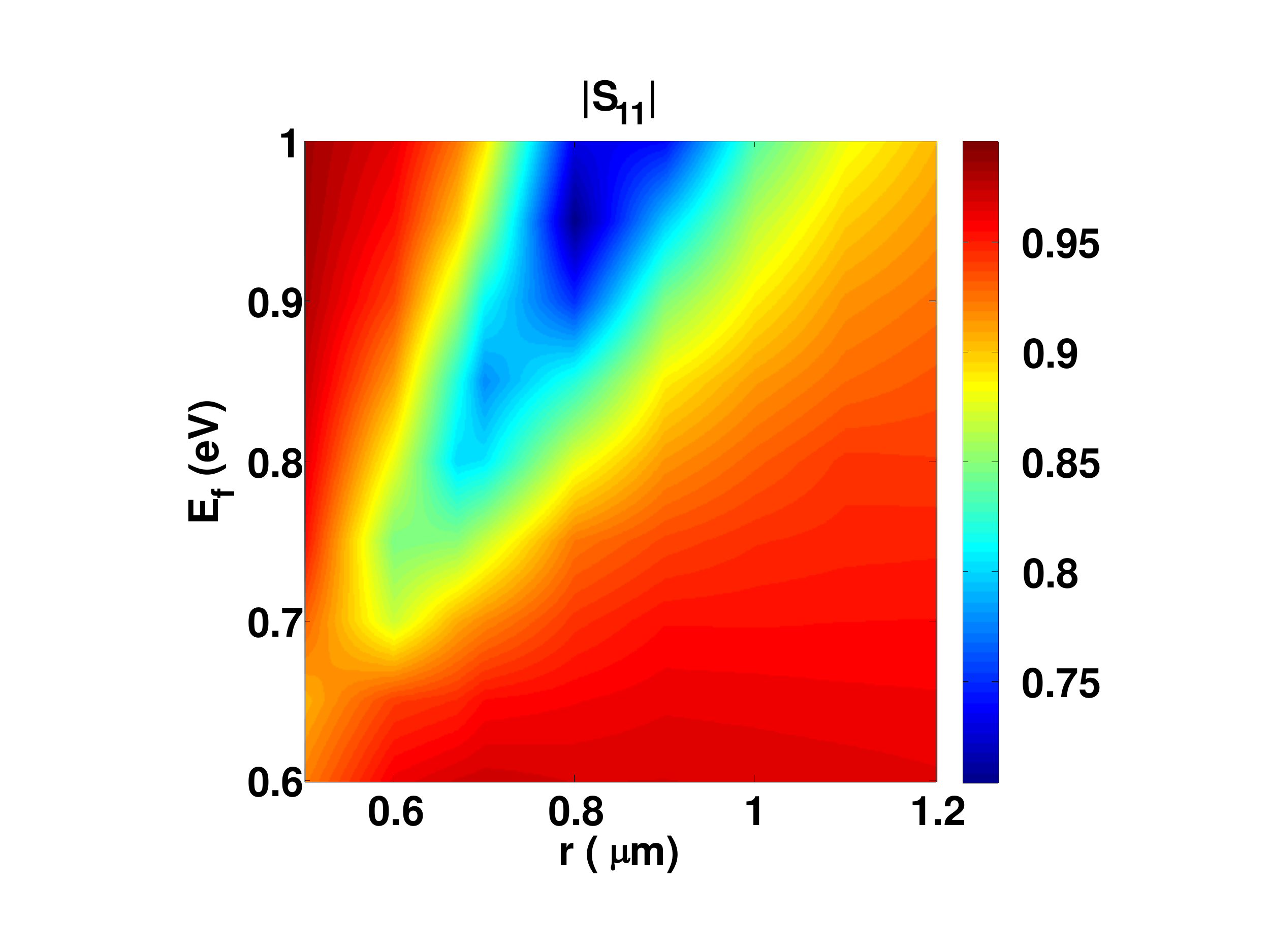}
	}
	\hspace{1cm}
	\subfloat[\label{2b}]{%
		\includegraphics[width=0.3\textwidth]{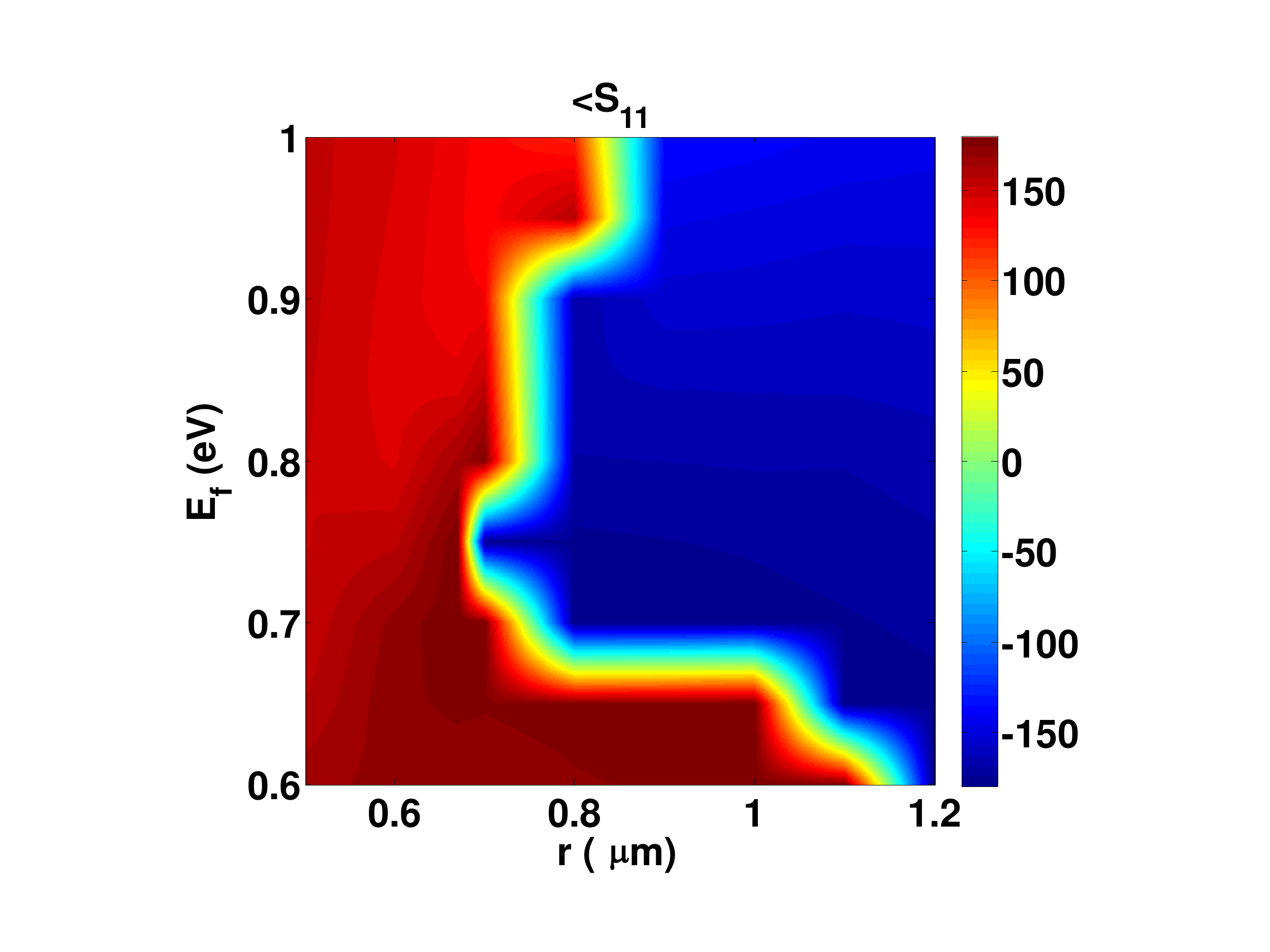}
	}
	\caption{ (a) Amplitude, and (b) phase of the reflection coefficient of a unit cell of the proposed hybrid metasurface  versus the Fermi level of the graphene patch $E_f$, and the radius of the hybrid nano-disk r.}
	\label{fig2}
\end{figure}
\section{Structure Design}

The schematic of the suggested hybrid plasmonic-dielectric is shown in Fig. \ref{fig1}. As it is observed, the structure is composed of a periodic array of nano-disks deposited on top of a dielectric spacer, namely silicon dioxide, with the thickness of $t_s$. Each nano-disk, in turn, is composed of a layer of silicon, with the thickness of $t_{d}$, and a graphene patch, with the surface conductivity of $\sigma_g$ on the top. A metallic gold layer with the thickness of $t_g$ is also embedded at the bottom of the whole structure so as to enhance the interaction of light with the metasurfce \cite{11}. \\
\indent In the proposed structure, the silicon layer and the graphene patch take the role of dielectric and plasmonic parts of the metasurface, respectively. To clarify this, we express the surface conductivity of graphene $\sigma_g$ in terms of its Fermi level $E_f$ and scattering rate $\Gamma$ as \cite{17,29,34}
{\setlength\arraycolsep{2pt}
	\begin{eqnarray}\label{Eq1}
	\sigma_{g}&&=i\frac{e^{2}K_{B}T}{\pi \hbar^{2} (\omega+2i\Gamma)}[\frac{E_f}{K_{B}T}+2\ln(\exp(-\frac{E_f}{K_{B}T})+1)]\nonumber\\
	&&+i\frac{e^{2}}{4\pi \hbar^{2}}\ln[\frac{2|E_f|-\hbar(\omega+2i\Gamma)}{2|E_f|+\hbar(\omega+2i\Gamma)}],
	\end{eqnarray}}
in which $\hbar$, $K_{B}$, $T$, and $\omega$ are reduced Planck constant, Boltzman constant, temperature in Kelvin, and angular frequency,  respectively, and $i=\sqrt{-1}$. It can be deduced from Eq. (\ref{Eq1}) that if the frequency of operation is limited to few terahertz, the surface conductivity $\sigma_g$ becomes inductive enabling the patch to support surface plasmons \cite{17}. \\
\indent Another interesting feature of the surface conductivity of the graphene patch, given in Eq. (\ref{Eq1}), is that it can easily be tuned via a change in the Fermi level $E_f$ of the graphene. Such appealing feature has made graphene an exceptional material to design reconfigurable optics and electronics components \cite{18,19,20}. In the following, we also have exploited this unique feature to have a full control over the phase of the metasurface, which provides the opportunity to demonstrate desired optical functionalities.\\
\indent To design the proposed metasurface for a specific optical functionality, the characteristics of a single unit cell should first be investigated. Here, without loss of generality, the periodicity of the metasurface and the thicknesses $t_d$ and $t_s$ are considered to be $L=3~\mu {\rm m}$, $t_d=3~\mu {\rm m}$ and $t_s=0.05~\mu {\rm m}$. Moreover, we assume the operation frequency to be $f=10~{\rm THz}$. The operation frequency is chosen such that the mentioned criteria regarding the surface conductivity of the graphene patch is maintained, and consequently it can guide surface plasmons.
\begin{figure}[b]
	\centering
	\subfloat[\label{subfig-1:dummy}]{%
		\includegraphics[height=0.23\textwidth]{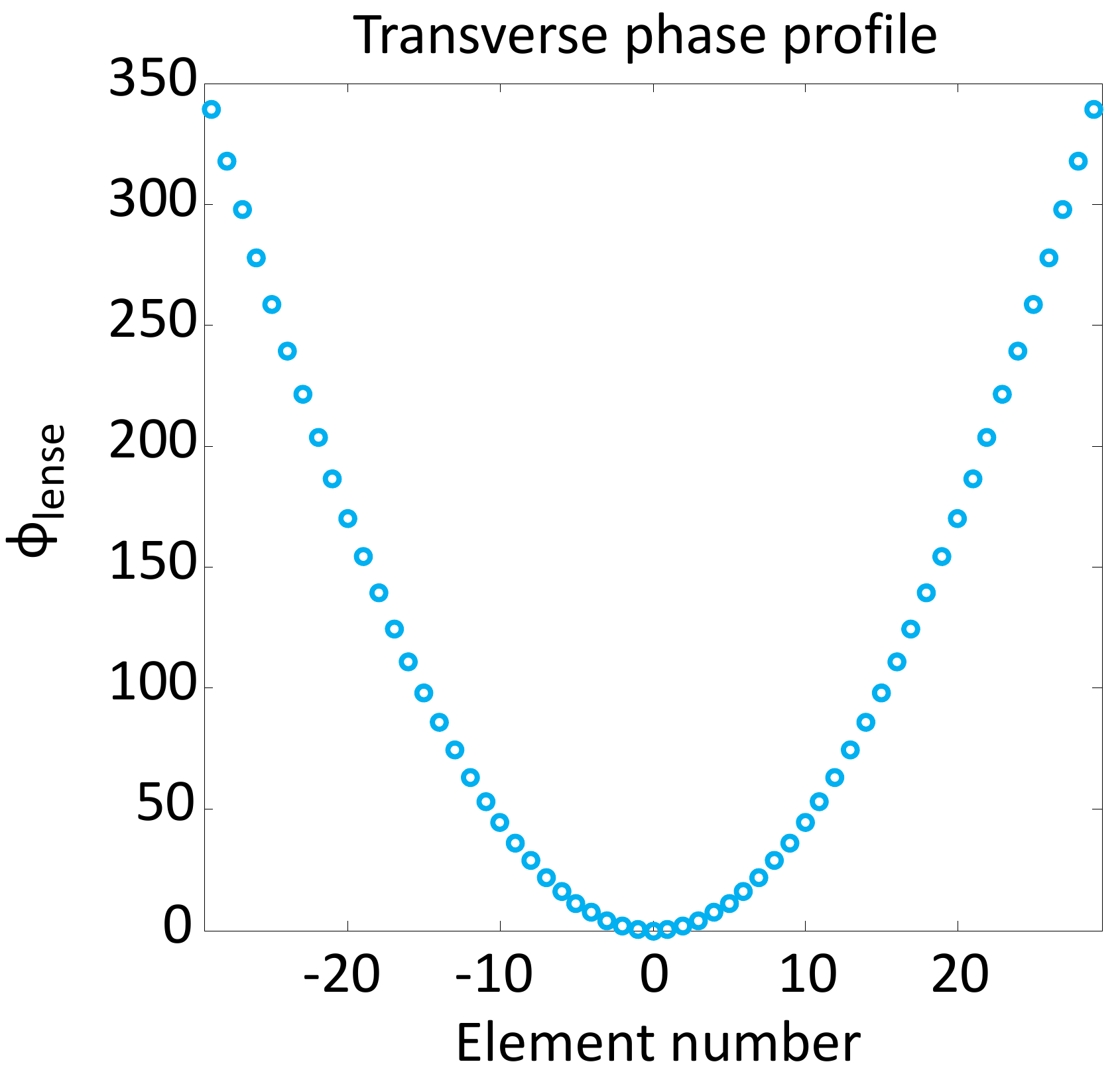}
	}
	\subfloat[\label{subfig-1:dummy}]{%
		\includegraphics[height=0.23\textwidth]{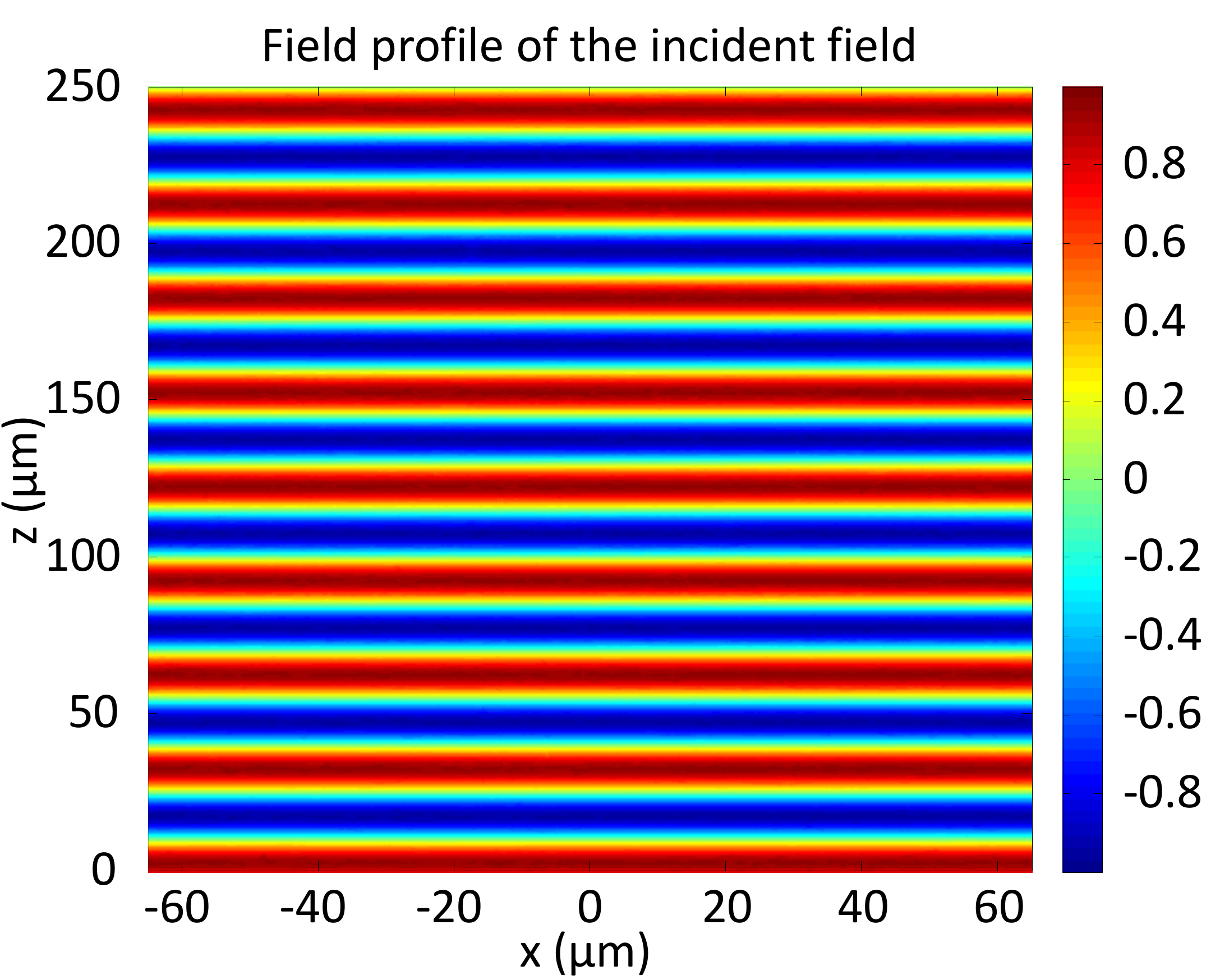}
	}
	\subfloat[\label{subfig-2:dummy}]{%
		\includegraphics[height=0.23\textwidth]{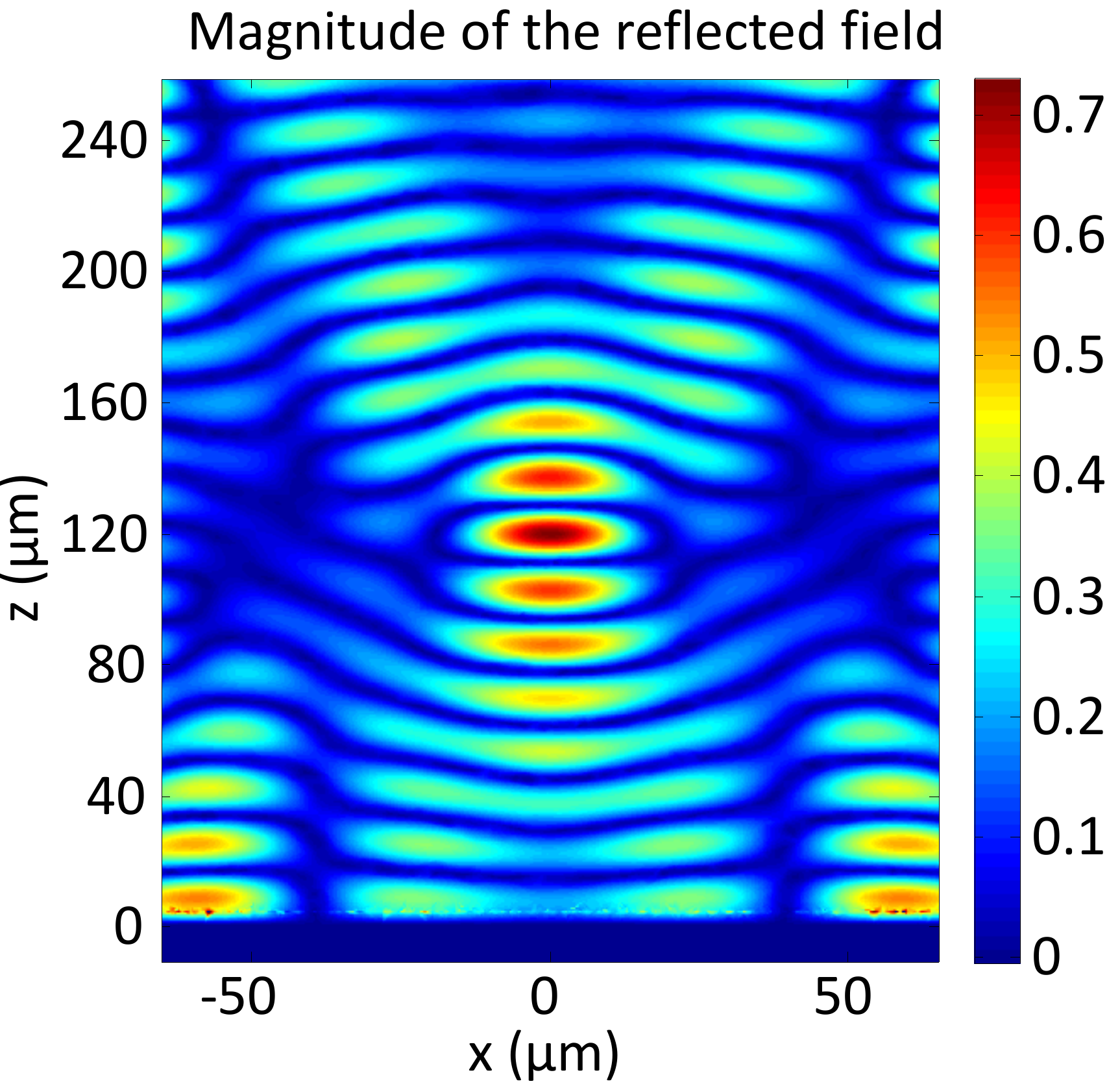}
	}
	\subfloat[\label{subfig-2:dummy}]{%
		\includegraphics[height=0.23\textwidth]{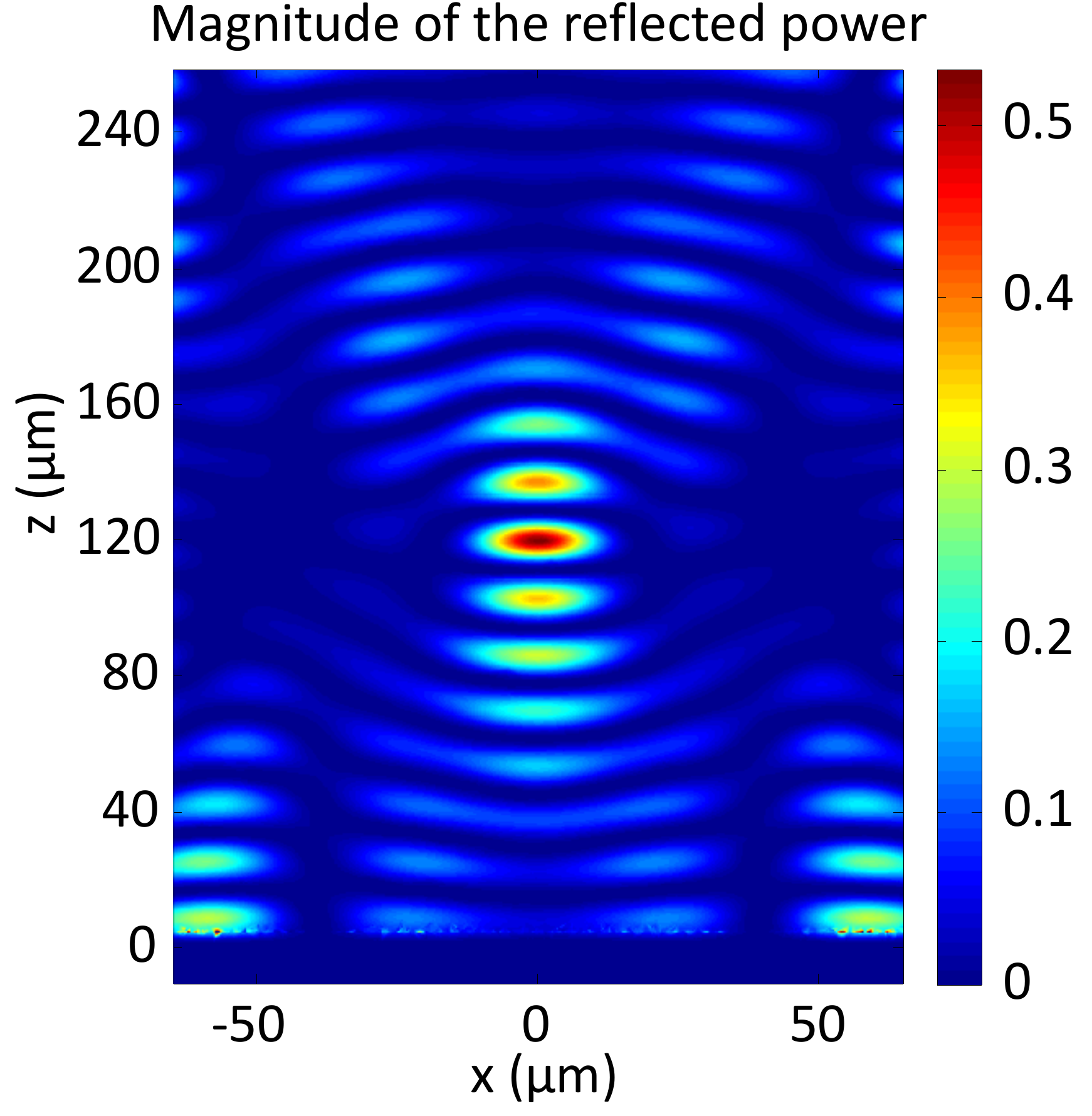}
	}
	\caption{(a) Required phase distribution to perform focusing at the focal length of $F=120~\mu {\rm m}$, (b) field profile of the incident field, (c) field profile of the corresponding reflected field, and (d) profile of the corresponding reflected power.}
	\label{fig3}
\end{figure}
\begin{figure}
	\centering
	\subfloat[\label{subfig-1:dummy}]{%
		\includegraphics[height=0.23\textwidth]{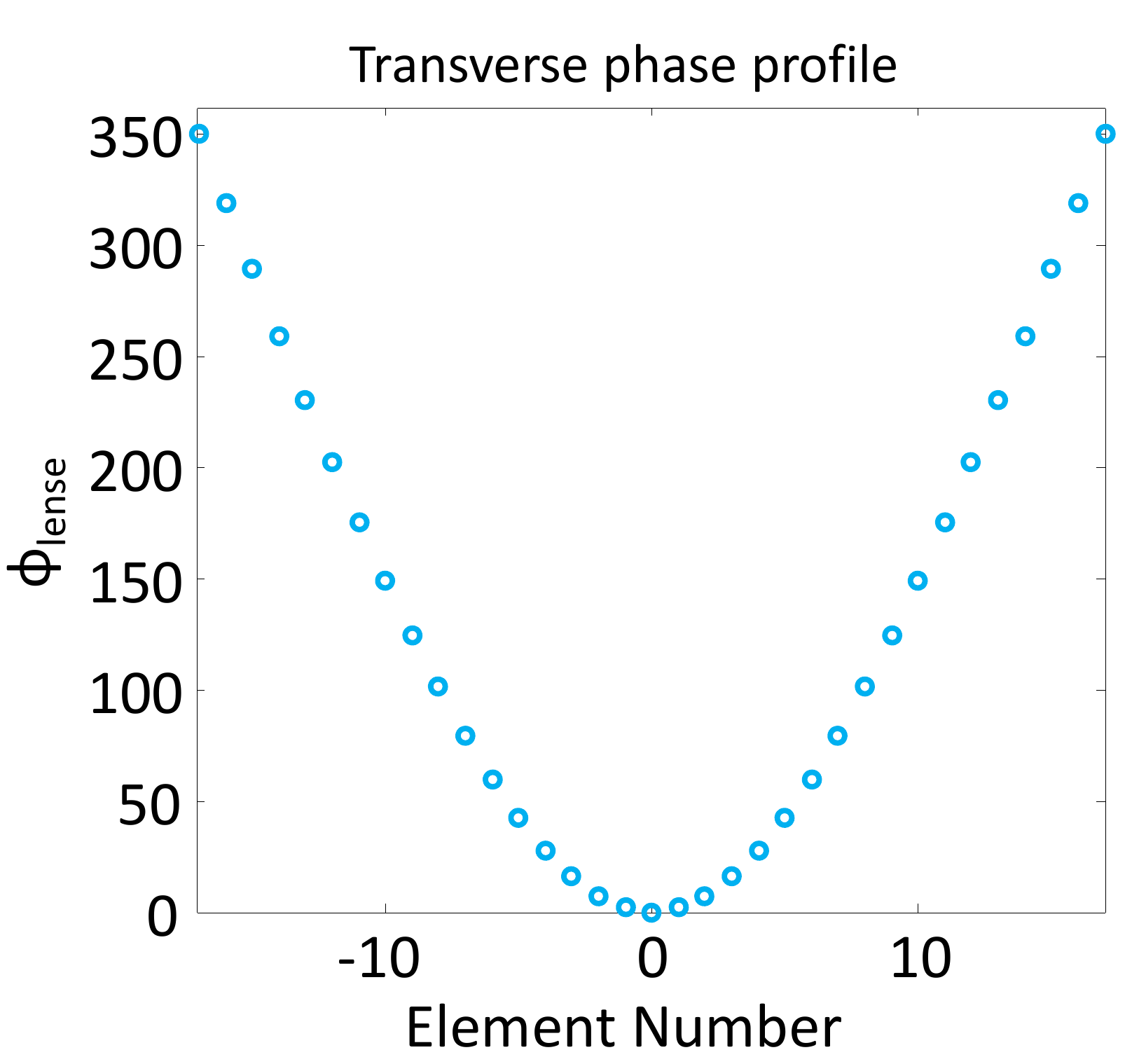}
	}
	\subfloat[\label{subfig-1:dummy}]{%
		\includegraphics[height=0.23\textwidth]{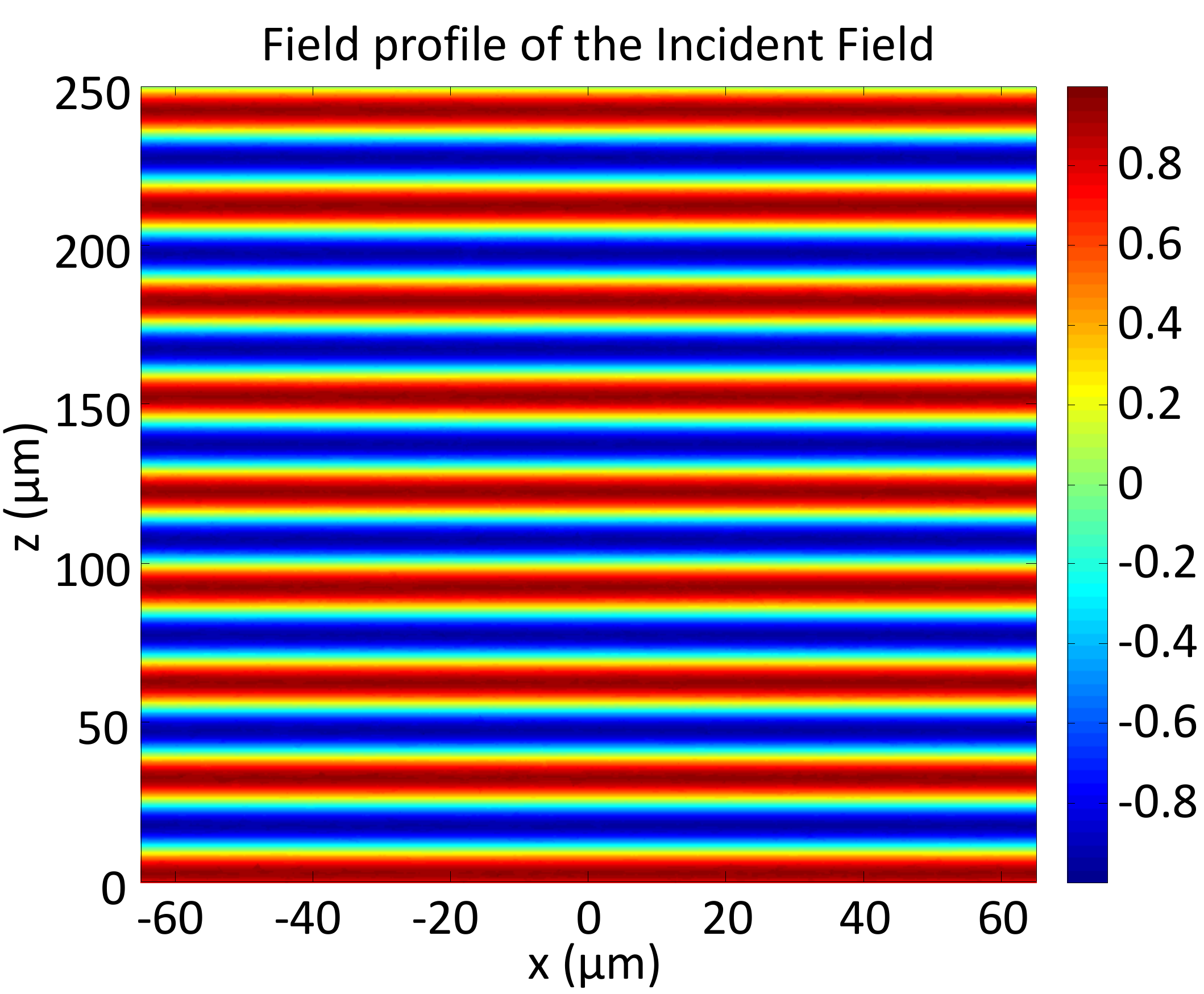}
	}
	\subfloat[\label{subfig-2:dummy}]{%
		\includegraphics[height=0.23\textwidth]{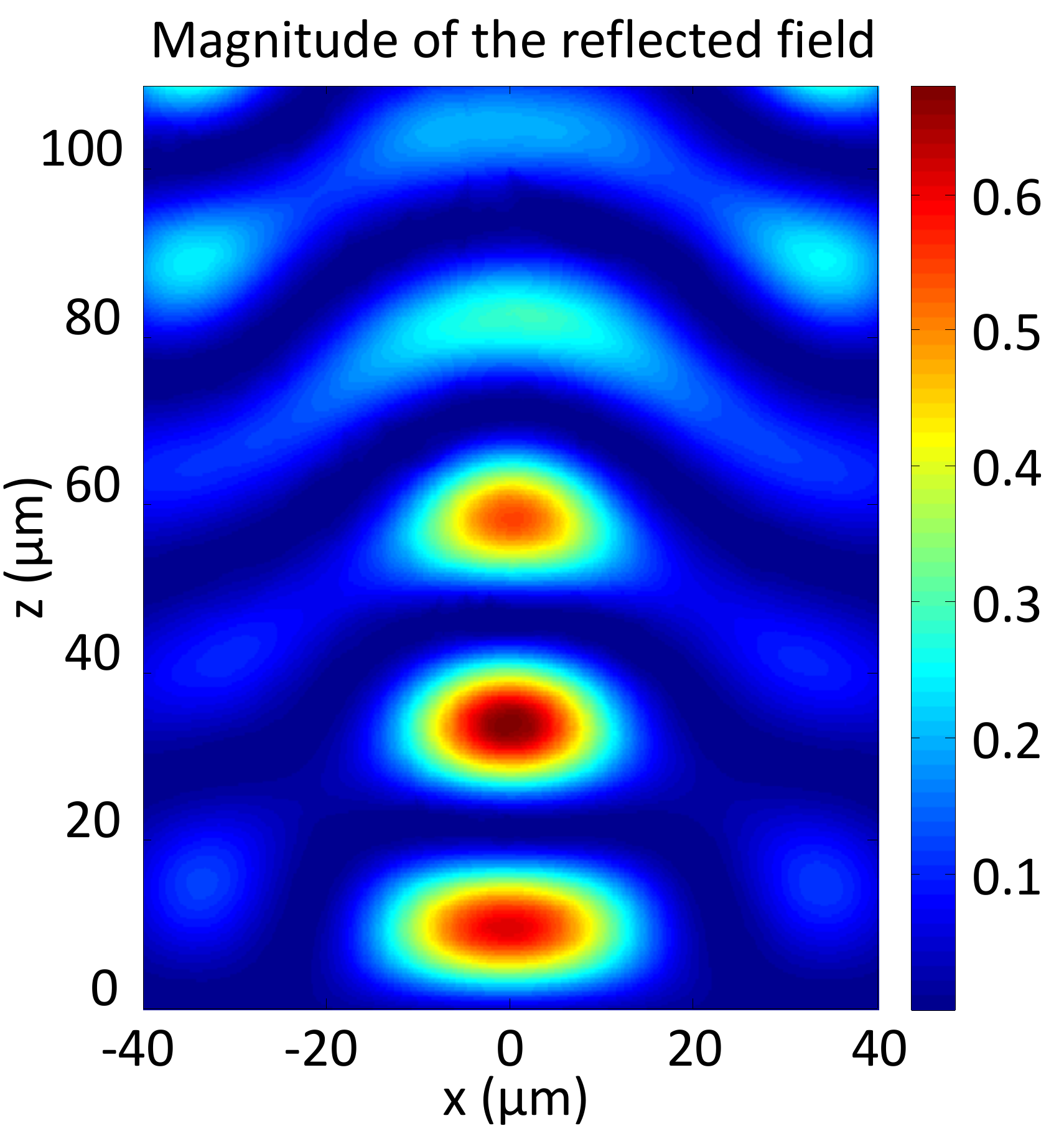}
	}
	\subfloat[\label{subfig-2:dummy}]{%
		\includegraphics[height=0.23\textwidth]{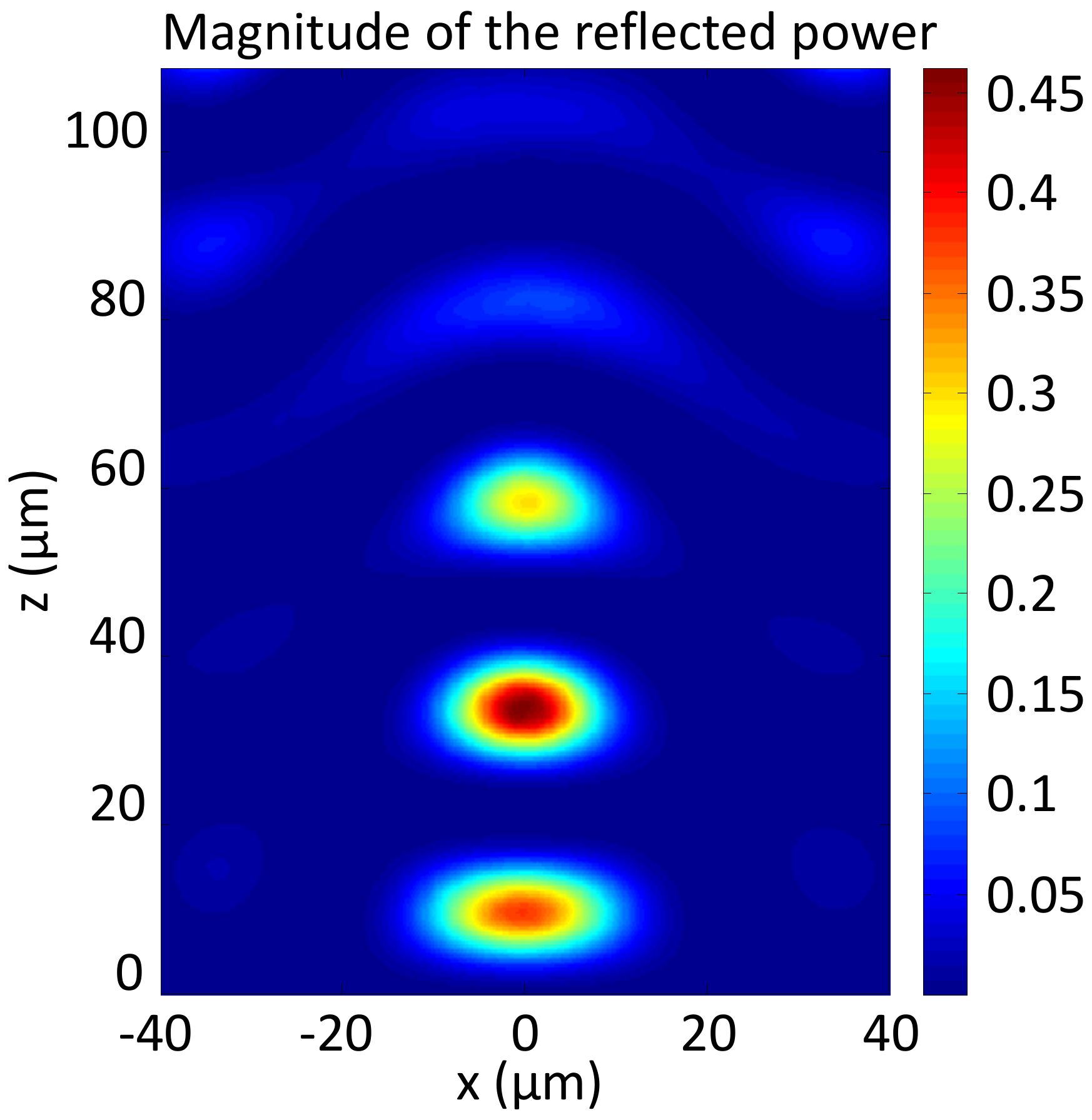}
	}
	\caption{(a) Required phase distribution to perform focusing at the focal length of $F=30~\mu {\rm m}$, (b) field profile of the incident field, (c) field profile of the corresponding reflected field, and (d) profile of the corresponding reflected power.}
	\label{fig4}	
\end{figure}
\indent figs. \ref{fig2}(a-b) show the amplitude and phase of the reflection coefficient of the unit cell versus the Fermi level of the graphene patch $E_f$, and the radius of the nano-disk $r$.  The results are obtained using a full-wave electromagnetic solver, namely HFSS. In HFSS, the graphene patches were modeled via impedance layers. The corresponding values of impedances were calculated simply by reversing the parameter $\sigma_g$ in Eq. (1).  As one can observe from Figs. \ref{fig2}(a-b), the magnitude of the reflection coefficient is higher than 0.75 and simultaneously its phase varies smoothly from $-\pi$  to $\pi$. The full control over the phase of the reflected field is due to the mentioned tunable conductivity of graphene. Moreover, the ohmic-lossless dielectric layer, i.e., silicon, keeps the amplitude of the reflection high. These features enable the proposed hybrid metasurface to perform high efficient optical functionalities.\\
\indent In order to perform a specific optical functionality, the phase of the corresponding reflected field to each element of the metasurface should be modulated properly. For instance, to perform focusing, the phase profile has to be modulated in a way that the reflected fields from single elements interfere with each other constructively. To maintain such condition, the phase profile should be of the form \cite{21}
\begin{equation}\label{5}
\phi_{\rm lens}(x)=\frac{2\pi}{\lambda}(\sqrt{F^{2}+x^{2}}-F)
\end{equation}
in which $F$ is the focal length and $\lambda$ is the wavelength of operation. Here, we consider a focal length of $F=4\lambda=120 ~\mu {\rm m}$ and $x = mL$, in which $L$ equals to the periodicity of the nano-disks and $m=0,\pm1,\pm2,...,\pm29,\pm30$. The required phase distribution is then calculated making use of  Eq. (2), and the result is depicted in Fig. \ref{fig3}(a). Based on the calculated phase profile of a single unit-cell of the metasurface in Fig. \ref{fig2}(b), we can then appropriately determine the radii and Fermi levels of nano-disks so that the phase of the reflected field of each element matches the phase distribution of Fig. \ref{fig3}(a). It should also be noted that the values of radii and the Fermi levels should be chosen in a way that the amplitude of reflection, shown in Fig. \ref{fig3}(a), becomes as high as possible. \\
\indent A plane wave with the field profile shown in Fig. \ref{fig3}(b) is then considered to perpendicularly be incident on the designed hybrid metasurface. The field profile of the corresponding reflected field is shown in Fig. \ref{fig3}(c). As it is observed, the reflected light is efficiently focused at the focal point of $F=120~\mu {\rm m}$, which is consistent with our expectation. The profile of the reflected power is also depicted in Fig. \ref{fig3}(d).\\
\indent In order to assess the tunability of the device, we then tune the values of the Fermi levels such that the focal length is adapted to $F=\lambda=30\mu {\rm m}$.  Similar results to those illustrated in Fig. \ref{fig3} are  then obtained and depicted in Figs. \ref{fig4}(a-d), respectively. As it is observed, the reflected field has been efficiently focused at the desired focal length. The opportunity to focus the reflected field in such a small focal length is as a result of the plasmonic behaviour of the metasurface \cite{3}. On the other hand, the overall reflectivity is high, which is because of the ohmic loss-less dielectric part of the metasurface. \\
\indent As another example, we demonstrate anomalous reflection functionality by means of the proposed metasurface. For bending the reflected field of an incident field towards a specific direction, $\theta_r$, the following linear phase profile has to be imprinted to the incident wave \cite{22}
\begin{equation}\label{5}
\phi_{\rm anomalous}(x)=\frac{2\pi}{\lambda}(\sin(\theta_r)-\sin(\theta_i))x,
\end{equation}
\begin{figure}
	\centering
	\subfloat[\label{subfig-1:dummy}]{%
		\includegraphics[height=0.29\textwidth]{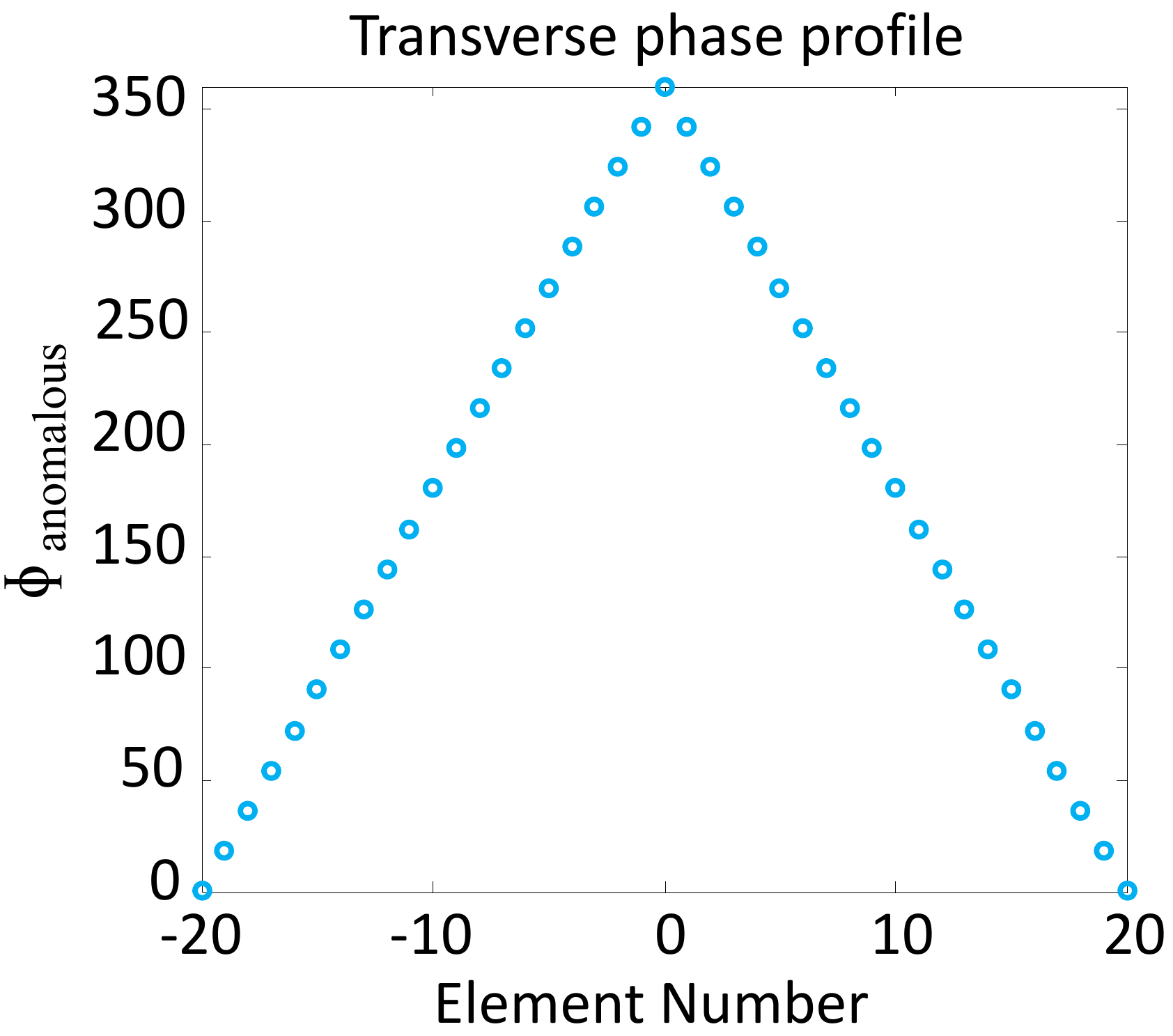}
	}
	\subfloat[\label{subfig-2:dummy}]{%
		\includegraphics[height=0.29\textwidth]{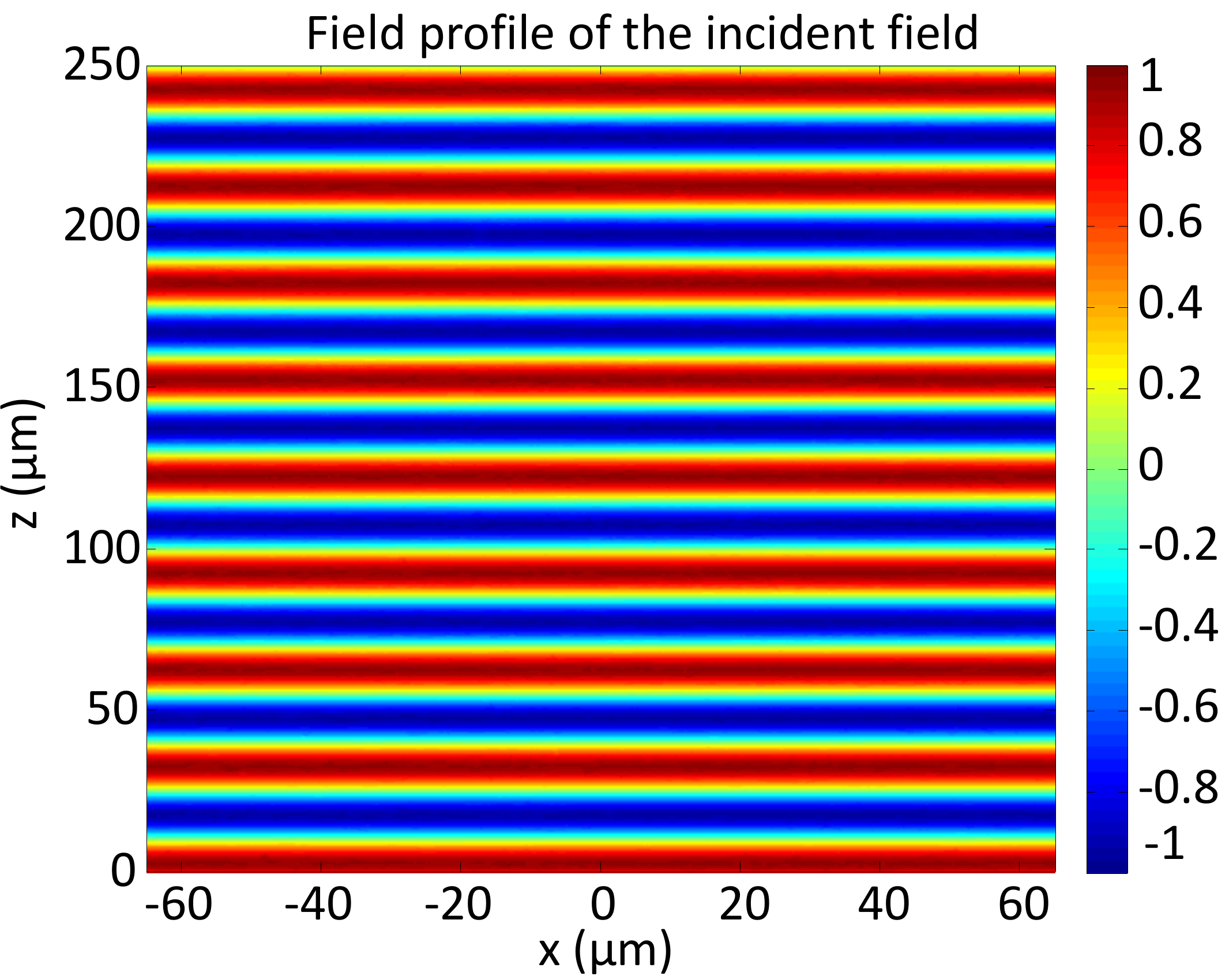}
	}
	\subfloat[\label{subfig-2:dummy}]{%
		\includegraphics[height=0.29\textwidth]{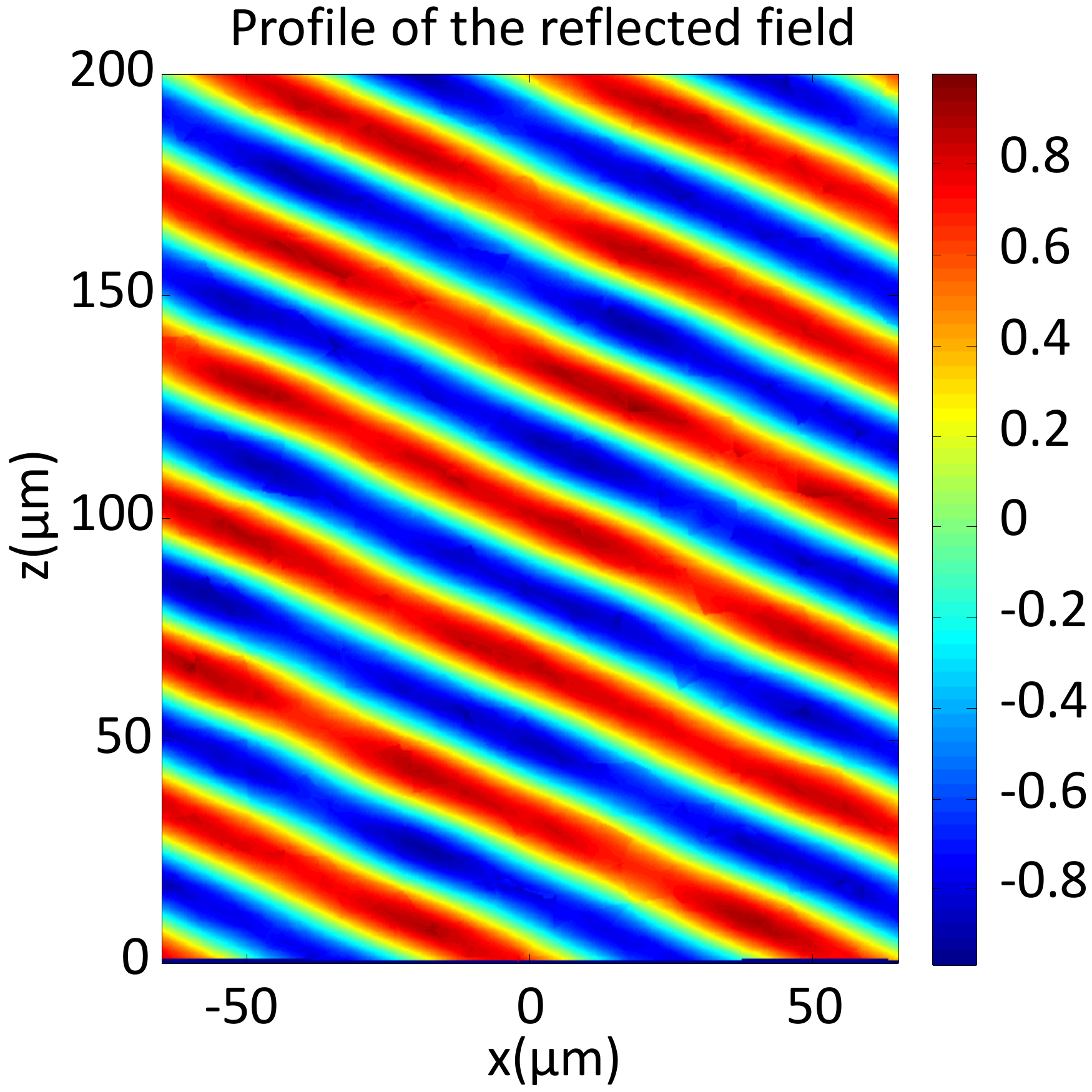}
	}
	\caption{(a) Required phase distribution to perform anomalous reflection for $\theta_i=0^{\circ}$ and $\theta_r=30^{\circ}$, (b) field profile of the incident field, and (c) field profile of the corresponding reflected field.}
	\label{fig5}
\end{figure}
where $\theta_i$ is the angle of incidence and $x = mL$, in which $L$ is the periodicity and $m$ is an integer. For the sake of simplicity, we consider the case in which the incident wave is perpendicularly incident onto the metasurface, i.e., $\theta_i=0^{\circ}$, and aim to bend the reflected wave toward the angle $\theta_r=30^{\circ}$.  Fig. \ref{fig5}(a) shows the required phase distribution $\phi_{\rm anomalous}$. Based on this figure, and by employing similar procedure presented previously, one can appropriately design the metasurface to bend the reflected wave in the desired angle, $\theta_r$.\\
\indent To evaluate the performance of the designed metasurface, a plane wave shown in Fig. \ref{fig5}(b) is considered to be incident on the designed structure, and then its associated reflected field is obtained and depicted in Fig. \ref{fig5}(c). As it is observed, the reflected field is bent to the desired angle $\theta_r=30^{\circ}$. \\
\indent It is worth noting that with respect to the previously proposed plasmonic metasurfaces, like those investigated in \cite{3,11,23}, our suggested hybrid metasurface has two main advantages: (1) compatibility to the current CMOS technology, and (2) low level of ohmic losses. On the other hand, in comparison with the previously proposed dielectric metasurfaces \cite{15,16}, the proposed structure has the advantage of tunability, which is directly related to the unique feature of graphene.

\section{Conclusions}
In this work, we proposed a novel hybrid plasmonic-dielectric metasurface for beam manipulation. The metasurface was composed of patterned hybrid graphene-silicon nano-disks deposited on a silica substrate. Two optical functionalities, namely focusing and anomalous reflection were demonstrated by means of the proposed hybrid  structure. The proposed hybrid metasurface had the advantages of previously proposed graphene-based  metasurfaces such as tunability, together with those of dielectric metasurfaces like  low ohmic lossees, and CMOS compability. One may also extend the proposed metasurface to perform various 3D optical functionalities.



\bibliography{sample}


\end{document}